\documentclass[aps,pre,twocolumn,amsmath,amssymb,floatfix,amsfonts,showkeys,superscriptaddress]{revtex4}
\usepackage{epsfig,color}
\usepackage{times}
\usepackage{amsfonts}
\usepackage{amssymb}
\usepackage{amsmath}
\usepackage{subfigure}
\usepackage{graphicx}
\usepackage{hyperref}
\usepackage{color} 
\usepackage{verbatim}
\usepackage{comment,xcolor}
\usepackage{textcomp}
\begin{document}
\title{Interactions mediated by a public good transiently increase cooperativity in growing
Pseudomonas putida metapopulations}

\author{Felix Becker}
\affiliation{Microbiology, Department Biology 1, Ludwig-Maximilians-Universität Munich,
Grosshaderner Strasse 2-4, Martinsried, Germany}

\author{Karl Wienand}
\affiliation{Arnold Sommerfeld Center for Theoretical Physics, Department of Physics, 
Ludwig-Maximilians-Universit\"at M\"unchen, Theresienstrasse 37, 80333 M\"unchen, Germany}

\author{Matthias Lechner}
\affiliation{Arnold Sommerfeld Center for Theoretical Physics, Department of Physics, 
Ludwig-Maximilians-Universit\"at M\"unchen, Theresienstrasse 37, 80333 M\"unchen, Germany}

\author{Erwin Frey}
\affiliation{Arnold Sommerfeld Center for Theoretical Physics, Department of Physics, 
Ludwig-Maximilians-Universit\"at M\"unchen, Theresienstrasse 37, 80333 M\"unchen, Germany}

\author{Heinrich Jung}
\affiliation{Microbiology, Department Biology 1, Ludwig-Maximilians-Universität Munich,
Grosshaderner Strasse 2-4, Martinsried, Germany}

\begin{abstract}
Bacterial communities have rich social lives. A well-established interaction involves the
exchange of a public good in \textit{Pseudomonas} populations, where the iron-scavenging
compound pyoverdine, synthesized by some cells, is shared with the rest. Pyoverdine thus
mediates interactions between producers and non-producers and can constitute a public good.
This interaction is often used to test game theoretical predictions on the ``social dilemma'' of
producers. Such an approach, however, underestimates the impact of specific properties of the
public good, for example consequences of its accumulation in the environment. Here, we
experimentally quantify costs and benefits of pyoverdine production in a specific
environment, and build a model of population dynamics that explicitly accounts for the
changing significance of accumulating pyoverdine as chemical mediator of social
interactions. The model predicts that, in an ensemble of growing populations
(metapopulation) with different initial producer fractions (and consequently pyoverdine
contents), the global producer fraction initially increases. Because the benefit of pyoverdine
declines at saturating concentrations, the increase need only be transient. Confirmed by
experiments on metapopulations, our results show how a changing benefit of a public good
can shape social interactions in a bacterial population.
\end{abstract}


\maketitle

\section{Introduction}
\label{intro}

Bacteria have complex social lives: they communicate with each other and with other
organisms, form tight communities in biofilms, exhibit division of labor, compete, and
cooperate \cite{1,2,3,4,5,6,7}. They also produce and exchange public goods. Public goods are chemical
substances that are synthesized by some individuals (known as \textit{producers} or \textit{cooperators}) and
are then shared evenly among the whole population, including cells that did not contribute to
their production \cite{8,9,10}. Such social interactions can also influence population dynamics, as
exemplified in the context of metapopulations \cite{11,12,13,14,15,16,17}. Metapopulations consist of several
subpopulations. The subpopulations may grow independently for a time, then merge into a
single pool that later splits again, restarting the cycle. This ecological system, which mimics
some bacterial life-cycles \cite{18,19}, also dramatically impacts the population's internal dynamics.
To mathematically analyze the effects of social interactions, they can be framed in terms of
game theoretical models \cite{20,21,22,23,24} for instance, the prisoner's dilemma, in the case of the
exchange of public goods \cite{25,26,27,28} – or formulated in terms of inclusive fitness models \cite{29,30,31}. These
approaches underestimate the impact on the social interaction of specific properties and
mechanisms of action of the public good in question, mostly to simplify the mathematical
description. Previous investigations have shown that, for example, phenomena like public
good diffusion \cite{32,33,34}, interference of different public goods with each other \cite{35}, the regulatory
nature of public good production \cite{36}, or its function in inter-species competition \cite{37} may affect
strain competition. The shortcomings of game-theoretical models in studying the evolution of
cooperation can be overcome by systems biology modeling approaches \cite{34,38,39}.

\begin{figure*}[htb]
\centerline{
\includegraphics{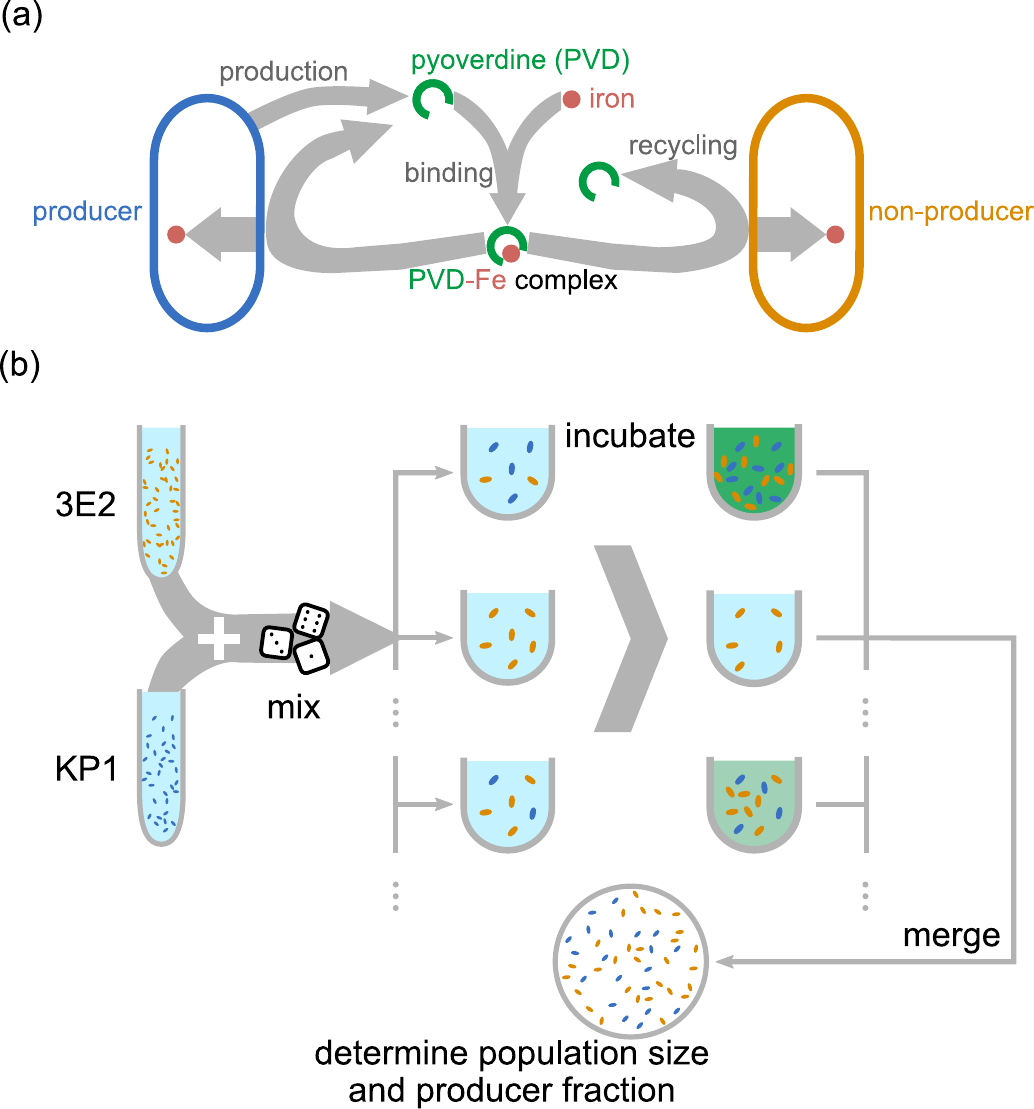}
}
\caption{Outline of PVD-mediated interactions and experimental setting. (a) Outline of the social
interaction. Producers (blue) secrete pyoverdine (PVD, green) into the environment, where it binds
iron (red). The resulting Fe-PVD complex is transported into the periplasm of both producers and non-
producers. Iron is reduced and incorporated into cells, while PVD is transported back into the \cite{44,45,46}
environment to scavenge additional ferric ions
. (b) Metapopulation growth setting. We initiate a
metapopulation by mixing producers and non-producers in random proportions and inoculating the
individual populations, which grow independently. At given time points $t$, we take samples from each
population, and merge them to determine the average population size and the global producer fraction
of the metapopulation.}
\label{fig:Fig1}
\end{figure*}

In this work, we directly quantify a social interaction mediated by a public good.
Thus, we adopt a systems biology approach, rather than a more reductive game-theoretical
one. We focus on the dissemination of iron-scavenging pyoverdine (PVD) in a
metapopulation of fluorescent \textit{Pseudomonas putida}, and study how its biological function
determines the population dynamics.
In this well-established, native model system, cells secrete PVD into the environment
to facilitate iron uptake when the metal becomes scarce \cite{29,40,41,42,43,44}. PVD binds to ferric iron and is
then actively transported into the periplasm. There, the iron is reduced, released and
transported across the plasma membrane, while PVD is secreted back into the environment \cite{44,45,46}. Figure \ref{fig:Fig1} outlines the PVD-mediated interaction between producer and non-producer cells
and the metapopulation set-up we use to study its effects on population dynamics. In the
following, we show, both experimentally and in computer simulations, that the global fraction
of producer cells across a metapopulation increases during growth, but only transiently. This
effect hinges on the specifics of PVD biochemistry, which elude a game-theoretical analysis.
Thus, our study shows that the specific features of the public good considered are the key
determinant of the outcome of the social interaction. Our experiments employ a well-defined
system, with a constitutive producer and a non-producer strain. The simulations use a
mathematical model based on quantitative measurements of PVD’s costs and benefits, as well
as its behavior as an accumulating public good. For appropriate values of the parameters, the
theoretical results match those of experiments with\textit{ P. putida} metapopulations.

\section{Results}
\subsection{Characterization of the model system}
To investigate the social role of public goods, we chose the soil bacterium \textit{P. putida} KT2440
as a model system. This is a well-defined system in which, as sketched in Fig. \ref{fig:Fig1}a, a single
public good mediates all cell-cell interactions. \textit{P. putida} KT2440 synthesizes a single type of
siderophore – a pyoverdine (PVD) molecule \cite{47}– and does not produce 2-heptyl-3-hydroxy-4-
quinolone or other known quorum-sensing molecules that might otherwise interfere with the
social interaction \cite{48,49,50}.

\begin{figure}[htb]
\centerline{
\includegraphics[width=\linewidth]{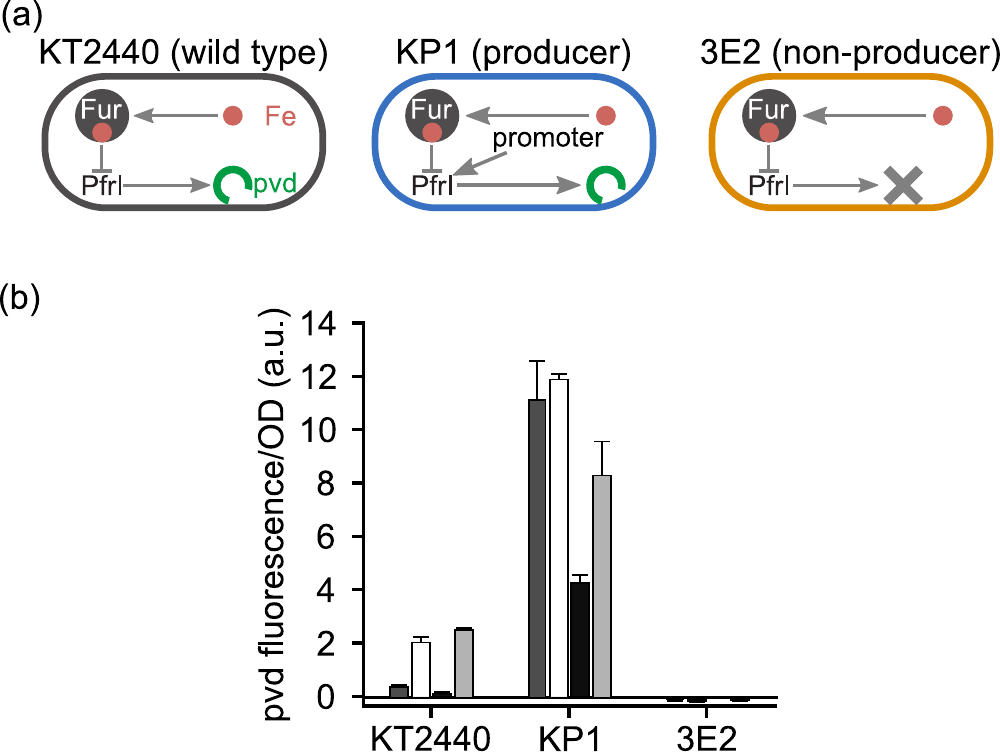}
}
\caption{Characterization of the strains. (a) Sketch of each strain's regulatory system. In the wild-type
\textit{P. putida} KT2440 (gray), the ferric uptake regulator Fur binds iron and represses the expression of the
pfrI gene necessary for PVD synthesis. The constitutive producer strain KP1 (blue) carries an
additional copy of the \textit{pfrI} gene controlled by a constitutive promoter. The non-producer strain 3E2
(orange) has an inactivated non-ribosomal peptide synthetase gene, which prevents PVD synthesis. (b)
Average PVD production per cell by the wild-type and strains KP1 and 3E2 after 8 h of cultivation.
The darker the columns, the more abundant is the iron in the medium. Dark gray columns represent
moderate iron availability conditions (KB medium without additions); white columns represent
extreme iron limitation (KB/1 mM DP); black columns represent iron-replete conditions (KB/100 $\mu$M
FeCl$_3$); light gray columns represent iron-limiting conditions (KB/100 $\mu$M FeCl$_3$ /1 mM DP). KP1
produces PVD under all conditions (albeit with different yields), 3E2 never produces the siderophore,
and the wild-type adapts its rate of synthesis to iron availability.}
\label{fig:Fig2}
\end{figure}

Wild-type \textit{P. putida} KT2440 controls PVD production through a complex regulatory
network. As shown in Fig. \ref{fig:Fig2}a, the central element of the network is the ferric uptake regulator
(Fur) protein, which binds iron and, among other things, down-regulates expression of the
iron starvation sigma factor \textit{pfrI} \cite{51,52,53}, which in turn directs the transcription of PVD synthesis
genes. As a consequence, siderophore production continually adapts to the availability of
iron \cite{47,52}. This regulation, however, obscures the costs of PVD production, as it also affects
other processes. We therefore circumvented it by generating a \textit{P. putida} strain, called KP1,
which constitutively produces PVD. KP1 carries a copy of the \textit{pfrI} gene controlled by theconstitutive promoter P$_{A1/04/03}$ \cite{54} at the attTn7 site in the KT2440 genome. As the non-
producer, we used strain 3E2, which carries an inactivated non-ribosomal peptide synthetase
gene (pp4220) that inhibits PVD synthesis \cite{47}. The two strains were otherwise isogenic.

We characterized producer (KP1) and non-producer (3E2) strains by measuring their
average per-cell PVD production under different iron availabilities, and comparing the results
with those for the wild type (strain KT2440). We cultivated all three strains, separately, in KB
medium and KB supplemented with 100 μM FeCl$_3$ (for short, KB/100 $\mu$M FeCl$_3$), as well as
in the same two media supplemented with the chelator dipyridyl (DP, 1 mM) to reduce iron
availability. Using atom absorption spectroscopy, we determined an iron concentration in KB
of about 8 μM. Figure \ref{fig:Fig2}b shows the average amount of PVD produced per cell after 8 h of
growth (close to the end of exponential growth). The wild type partially represses production
of PVD under moderate iron availability (KB, dark gray bars), and ceases synthesis altogether
under high iron availability (KB/100 $\mu$M FeCl$_3$, black bars). Addition of DP reduces iron
availability and stimulates PVD production in both media (Fig. \ref{fig:Fig2}b, white bars: KB/1mM DP,
light gray bars: KB). In contrast, KP1 produces large amounts of PVD under all tested growth
conditions, and thus represents a constitutive PVD producer. The yield depends on conditions,
probably because the regulated copy of \textit{pfrI} is still present in the genome. 3E2, finally, never
synthesizes PVD, regardless of the conditions, and is a true non-producer, as previously
reported \cite{47}.

\subsection{Quantifying the social role of pyoverdine}
\begin{figure}[htb]
\centerline{
\includegraphics[width=\linewidth]{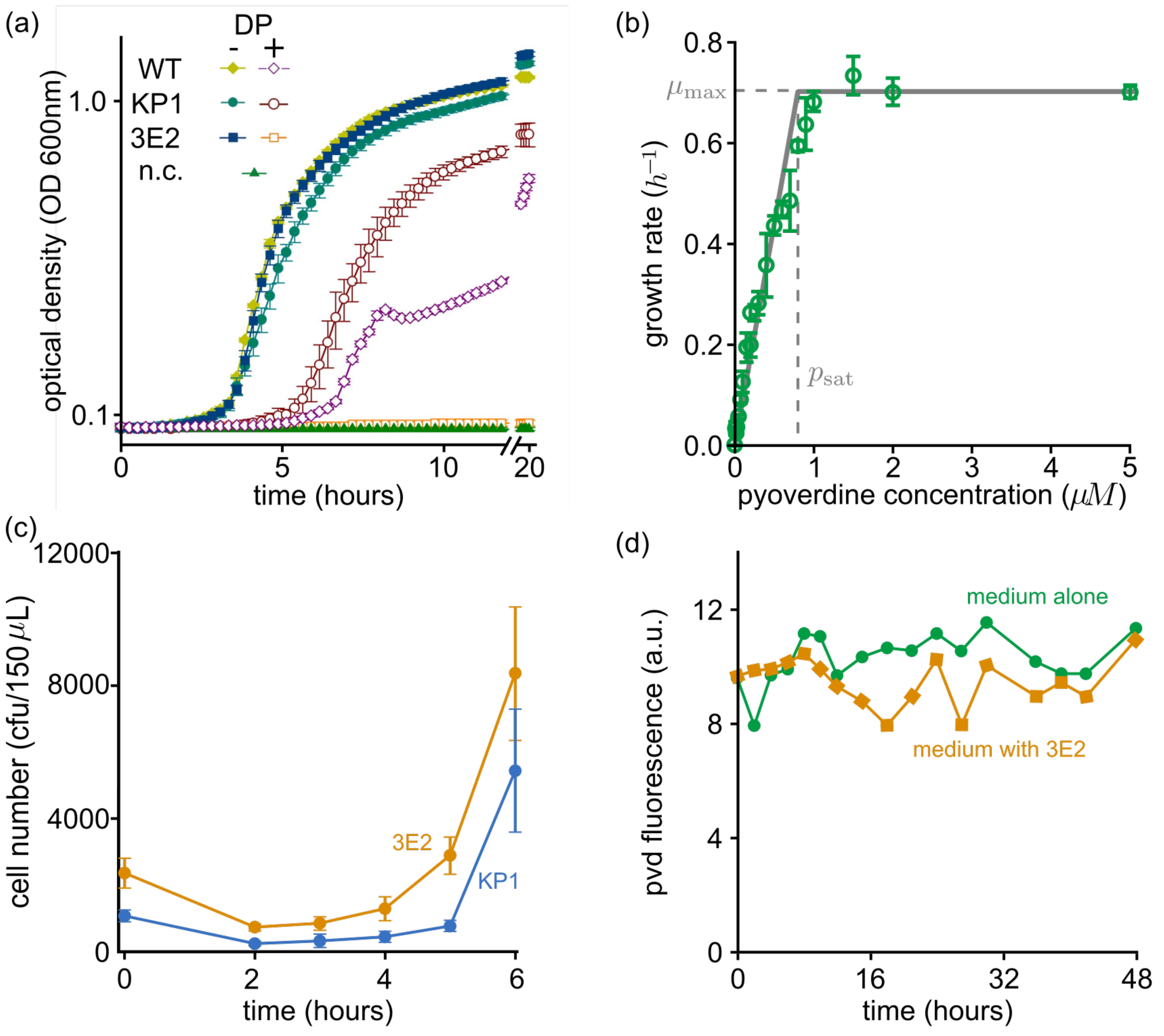}
}
\caption{Characterization of the social impact of pyoverdine (PVD) in terms of costs (a), benefits (b),
degree of sharing among cells (c) and stability (d). (a) In an environment with available iron (KB,
solid symbols), non-producer cells (strain) 3E2 grows as fast as the wild-type (WT), and faster than
the producer (strain KP1). Under extreme iron limitation (KB /1 mM DP, empty symbols), PVD is
needed for growth: KP1 and WT grow, whereas 3E2 does not (mean values and standard deviations
were calculated from six measurements). (b) Green dots represent the growth rate $\mu$ of 3E2 cultures,
measured under extreme iron limitation (KB/1 mM DP) in the presence of the indicated concentrations
of added PVD (error bars are standard deviation over four replicates). The solid gray line represents
the growth rate calculated using equation (\ref{eq:2}) (maximal growth rate $\mu_\textnormal{max}$ and the saturation
concentration p sat fitted to the experimental data: $\mu_\textnormal{max}$=0.878, $p_\textnormal{sat}$=0.8). (c) Early growth of KP1-
3E2 co-cultures under extreme iron limitation (KB/1 mM DP). Shown are mean and SD of eight
independent experiments. Since 3E2 needs PVD to grow (see panel b), this result indicates that PVD
is shared between the strains. (d) Fluorescence of PVD in KB medium and in the presence of non-producer bacteria.}
\label{fig:Fig3}
\end{figure}

Having established how each strain behaves, we quantified the impact of PVD on
population dynamics. Specifically, we wanted to determine the metabolic load of PVD
production, its impact on growth, its stability, and how evenly it is shared with other cells.
We assessed the impact of PVD production on growth by comparing the growth rates
of strains KP1 and 3E2 under iron-rich conditions (KB). As shown in Fig. \ref{fig:Fig2}b, neither 3E2 nor
the wild-type produces substantial amounts of PVD under these conditions, and the solid
symbols in Fig. \ref{fig:Fig3}a show that both strains grow at about the same rate. KP1, on the other hand,
produces PVD and grows more slowly. The data in Supplementary Table S1 allow us to
quantify this difference in growth rate. Depending on the conditions, KP1’s growth rate is 3-
10\% lower than that of strain 3E2. For example, the difference is minimized (1.03-fold) when
the medium is replaced (in a 24-well plate format) every hour, whereas the largest difference
(1.10-fold) is observed in batch cultures (96-well plate format). This suggests that factors
other than iron level \textit{per se}, such as nutrients and oxygen availability, modify the metabolic
impact of PVD production.
The empty symbols in Fig. \ref{fig:Fig3}a illustrate the growth of the strains under extreme iron
limitation (KB/1 mM DP). In these conditions, PVD is indispensable for iron uptake, and only
producing strains – KP1 and the wild-type – can grow at all. Less restrictive conditions
(KB/100 $\mu$M FeCl$_3$ and KB/100 $\mu$M FeCl$_3$/1 mM DP) produce qualitatively similar results
(see Supplementary Fig. S1). 3E2, if cultivated alone, cannot grow unless the medium is
supplemented with PVD isolated from a producer culture. Figure \ref{fig:Fig3}b shows the maximal
growth rate of 3E2 under these conditions as a function of the concentration of added PVD.
For values lower than about 1 $\mu$M, the growth rate increases almost linearly with PVD
concentration, then sharply levels off. Higher PVD concentrations do not further stimulate
growth – which is consistent with observations of iron saturation in other bacterial
systems \cite{55,56}.
This saturating behavior, we argue, stems directly from PVD’s ability to bind iron and
make it available to cells. Because PVD has an extremely high affinity for iron ($10^24$ M$^-1$ for
Fe$^{3+}$ at pH 7 \cite{57}), we can assume that each PVD molecule immediately binds an iron ion.
Therefore, the PVD concentration $p$ is equivalent to that of PVD-Fe complexes, and
represents the concentration of iron accessible to cells (this may not hold if the level of PVD
exceeds that of the iron available, but we expect this extreme case to arise only after the
exponential growth phase in our setting, if ever). Each cell, then, incorporates iron ions at a
constant rate $k\centerdot p$ which is proportional to the PVD concentration $p$. Moreover, cells try to
maintain a constant internal iron concentration $Fe_\textnormal{in}$ and reproduce at a PVD-dependent rate
$\mu(p)$ when growth is limited by iron availability. If we also assume that the cell volume just
before division is twice the volume $V_0$ of a newborn cell, we find that the growth rate is
proportional to $p$ (see Supplementary Note):

\begin{equation}
\mu(p)=\frac{k}{Fe_\textnormal{in} V(0)}p\,.
\label{eq:1}
\end{equation}

For PVD concentrations above 1 $\mu$M, however, some other factor limits growth. Cells cannot
further increase $\mu(p)$ , regardless of the PVD concentration, and the benefit of PVD saturates.
In summary, there is a limit PVD concentration $p_\textnormal{sat}$ ($\sim$1 μM), below which the growth rate is
proportional to the PVD concentration, following equation (\ref{eq:1}). Above $p_\textnormal{sat}$, the growth rate is
a constant $\mu_\textnormal{max}$, whose value depends on the culture conditions. In mathematical terms,

\begin{equation}
\mu(p)=\mu_\textnormal{max}\min\left(\frac{p}{p_\textnormal{sat}},1\right)=\begin{cases}
\frac{\mu_\textnormal{max}}{p_\textnormal{sat}}, & p<p_\textnormal{sat}\\
\mu_\textnormal{max} & p\geq p_\textnormal{sat}
\end{cases}
\,.
\label{eq:2}
\end{equation}

The gray curve in Fig. \ref{fig:Fig3}b shows the function described in equation (\ref{eq:2}). Fitting the values for
the parameters $p_\textnormal{sat}$ and $\mu_\textnormal{max}$, the curve closely resembles the experimental results, validating our argument.

A central question in determining the social role of PVD is whether cells share the
molecule with other cells, and thus also its benefit, or keep it to themselves. In other words, to
what extent is PVD a \textit{public} good? Fig. \ref{fig:Fig3}c shows the early stages of growth of a mixed
population of KP1 and 3E2 (initial size $5\times10^3$ cells per 150 $\mu$L) under extreme iron
limitation (KB/1 mM DP). After a lag phase of about 2 h, both strains begin to grow. Since
3E2 needs PVD to grow in these conditions (see above and Supplementary Fig. S1), we
conclude that both strains receive the benefit at the same time, and neither has preferential
access to it. In our experiments, then, PVD behaves as a truly public good. Consequently,
populations that start with a higher producer fraction $x_0$ have more PVD available, and grow faster than populations with low $x_0$ values (as shown in Supplementary Figs. S1 and S2).
PVD is also very stable. Figure \ref{fig:Fig3}d shows the fluorescence yield of PVD over 48 h in
KB medium alone (green line). The value fluctuates around a constant average, indicating that
PVD does not spontaneously degrade – at least not appreciably – within the time scales of our
experiments. The orange line in Fig. \ref{fig:Fig3}d represents a similar measurement, but in the presence
of non-producer cells. In this case also, fluorescence does not appreciably decay, so cells do
not seem to consume PVD during the interaction. This also means that, provided producers
are present, the public good accumulates in the environment once its synthesis has been
triggered.
Taken together, these observations characterize the social interaction as follows: (i)
Constitutive producers grow more slowly than non-producers (given equal PVD availability);
(ii) PVD acts as a public good, which is homogeneously shared among cells; (iii) once
produced, PVD persists: it is chemically and functionally stable, and cells recycle it rather
than consuming it; (iv) the public good drives the population dynamics, since PVD is
necessary for access to the iron required for growth.

\subsection{Modeling social and growth dynamics}
Based on the experimental results presented in the previous section, we formulated a set
of equations to describe the development of a single, well-mixed population of $c$ producers
and $f$ non-producers. The population dynamics follows a logistic growth, where the function
$\mu(p)$ from equation (\ref{eq:2}) determines the per-capita growth rate. For our experimental setup, we
estimate cells to incorporate only a minimal fraction of the available iron ($<3\%$, see
Supplementary Note), so the assumptions of equation (\ref{eq:2}) hold (and some other resource
determines the carrying capacity $K$). Although KP1 synthesizes PVD at condition-dependent
rates, we adopt a simplified description and model synthesis as occurring at a constant rate $\sigma$.
The produced PVD does not decay but accumulates in the medium. Finally, the costs of PVD
synthesis slow down the growth of producers by a factor $1 − s$ (where $s < 1$), compared to
non-producers. All in all, assuming the interaction between cells and PVD is fast, the
dynamics can be summarized in the following equations:
\begin{eqnarray}
\nonumber \frac{dc}{dt} &=& c\mu(p)(1-s)\left(1-\frac{c+f}{K}\right)\,,\label{eq:3}\\
\frac{df}{dt} &=& c\mu(p)\left(1-\frac{c+f}{K}\right)\,,\\
\nonumber \frac{dp}{dt} &=& \sigma c\,.
\end{eqnarray}
This set of equations mathematically describes the experimental facts, in terms of measurable
quantities. It is also different from a traditional game theoretical formulation, which would
require us to somehow define a payoff function.

To better highlight the key factors of the population dynamics, we rescale the
variables in equations (\ref{eq:3}). First, we measure population size in terms of the fraction of
resources used up, i.e., $n: = (c + f)/K$. This definition means that $K$ determines the scale of
population sizes, while n takes values between 0 and 1: as n approaches 1, the resources
become depleted, and cells enter a dormant state \cite{14}. Second, we consider the fraction
$x: = c/c + f$ of producers within each population, rather than their absolute number. Third,
we measure the amount of PVD in units of the saturation concentration, $v: = p/p_\textnormal{sat}$ (and
define $\mu(v) =\min(v, 1)$. Finally, measuring time in units of the minimal doubling time
$1/\mu:_\textnormal{max}$, equations (\ref{eq:3}) become
\begin{eqnarray}
\nonumber \frac{dn}{dt} &=& n(1-sx)(1-n)\mu(v)\,,\label{eq:4}\\
\frac{dx}{dt} &=& -sx(1-x)(1-n)\mu(v)\,,\\
\nonumber \frac{dp}{dt} &=& \alpha n x\,,
\end{eqnarray}
where $\alpha:=\frac{\sigma K}{p_\textnormal{sat}\mu_\textnormal{max}}$ is a dimensionless parameter. This parameter represents the rate at which
PVD benefit saturation sets in. Keeping other factors constant, the benefit saturates sooner if
production is faster (higher $\sigma$) and/or the number of total producers increases (higher $K$ and
thus larger populations). Conversely, if the saturating PVD concentration is higher (higher
$p_\textnormal{sat}$), or cell reproduction is faster (higher $\mu_\textnormal{max}$), populations can reach higher densities before
the benefit saturates. Generally speaking, the lower $\alpha$, the more advantageous producers are
for their population. For $\alpha\to 0$ for example, the reproduction time scale is shorter than that of
public good production. Therefore, the relatively scarce PVD strictly limits growth, PVD
saturation occurs only after many generations, and producer-rich populations outgrow
producer-poor communities for longer. At the other extreme, $\alpha\to\infty$ means that cells produce
PVD much faster than they grow. In this case, a handful of producers suffices to quickly reach
saturation levels of PVD. Whether they include few or many producers, all populations grow
at the same rate, which negates the advantage of higher producer fractions.

\begin{figure}[htb]
\centerline{
\includegraphics[width=\linewidth]{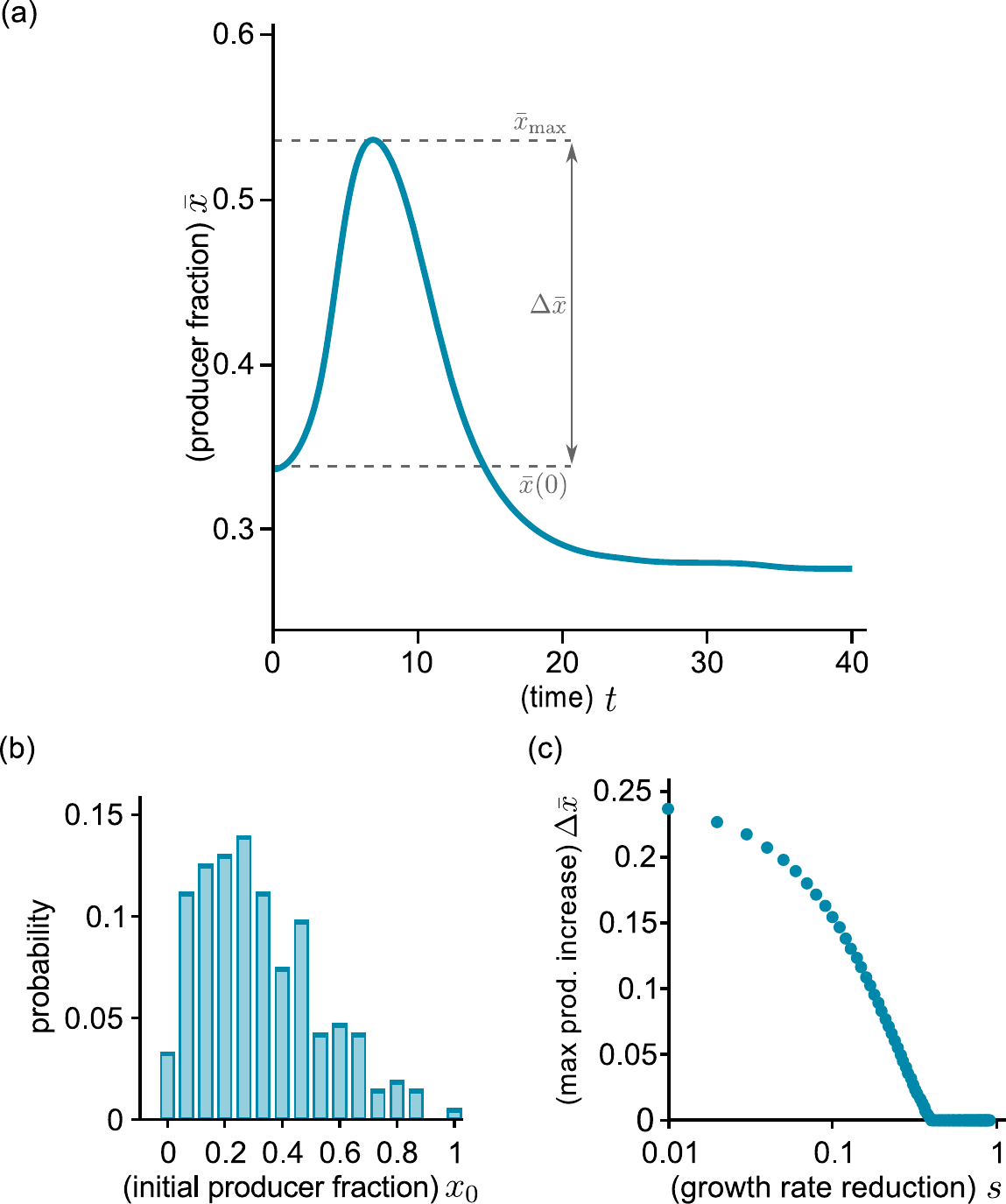}
}
\caption{Simulation results for the growth of a metapopulation. The time course of the global
producer fraction $\bar x$ (a) is computed by numerically solving equations (4) for a given distribution of
stochastic initial compositions (b) (parameter values: $\alpha=200$, $s=0.05$, initial size $n_0 = 10^{-3}$). The
producer fraction initially increases as populations with more producers begin to expand earlier (see
also Supplementary Movie S1). After reaching a maximum value $\bar x_\textnormal{max}$, the global producer fraction
decreases. (c) The maximal magnitude of the increase $\Delta\bar x = \bar x_\textnormal{max}-\bar x(0)$ decreases with stronger
growth reduction $s$ ($s$ between $0.01$ and $0.9$, other parameters identical to panel (a): for low $s$ it is
comparable to the initial producer fraction, while very low producer growth precludes any increase at
all.}
\label{fig:Fig4}
\end{figure}

We can also use equations (\ref{eq:4}) to describe a metapopulation of M independent
populations. To simulate this scenario, we solve equations (\ref{eq:4}) numerically for an ensemble of
stochastic initial conditions (using $M = 10^4$). We generate a stochastic distribution of initial
producer fractions $x_0$ – depicted in Fig. \ref{fig:Fig4}b – as implemented in the experiments (see Fig. \ref{fig:Fig5}a
and Materials and Methods). Because the experiments described here deal with relatively
large populations (starting with around $10^3$ - $10^4$ individuals, and expanding to between $10^6$
and $10^7$ cells), stochasticity in the initial size is low, and we initialize all populations in the
simulated ensemble with the same size $n_0= 10^{-3}$. Once the populations are formed, the
choice of $s$ and $\alpha$ completely determines the population dynamics.

During the simulations, we record the average size $\bar n = \frac{1}{M}\sum\limits_{i=1}^{M} n_i$
producer fraction $\bar x$ across the metapopulation
\begin{equation}
\bar x = \frac{\sum_i c_i}{\sum_i (c_i+f_i)}\,,
\label{eq: 5}
\end{equation}
where $n_i$ and $x_i$ are the size and producer fraction of each population $i$, respectively. Note that
this \textit{global} fraction of public-good producers (i.e., the percentage of producer cells in the
metapopulation) follows a different trajectory from the \textit{local} one $x_i$ (the fraction of producers
actually present in each of the component subpopulations). Specifically, while the latter
always decreases – because producers grow more slowly than non-producers – the former
can, in some cases, increase.

How x changes in time within a metapopulation, according to equations (\ref{eq:4}) (with
$\alpha = 200$, $s = 0.05$, $n_0 = 10^{-3}$, and compositions sampled from the
distribution in Fig. \ref{fig:Fig4}b, with average $\bar x (0) \simeq 0.33$) is shown in Fig. \ref{fig:Fig4}a; Supplementary Movie S1 shows the same
data, together with the evolution of the joint distribution of sizes $n_i$ and compositions $x_i$.
During early stages of growth, the more producers a population has, the quicker it accrues
PVD, and the faster it grows. Populations with higher producer fractions rapidly increase their
share in the metapopulation, driving up the global producer fraction $\bar x$. As time passes,
populations with fewer producers also accumulate enough PVD to grow significantly (while
the few with no producers never grow). Meanwhile, producer-rich populations have depleted
their resources and end growth. As a result, the rate of increase of $\bar x$ first slows, then reaches a
maximum $\bar x_\textnormal{max}$ and decreases again. Finally, once all populations have entered the dormant state, the global producer fraction stabilizes. Its ultimate value depends on the production cost
$s$ and, because all populations grow to the same size, it is lower or equal to the initial $\bar x(0)$.

The overall time course of $\bar x$ and $\bar n$ depends crucially on the choice of parameters,
which reflect the features of the bacterial strains, as well as the cultivation conditions. Fig. \ref{fig:Fig4}c,
for example, shows how changing the growth reduction s affects the magnitude of the
increase in global producer fraction $\Delta\bar x = \bar x_\textnormal{max}-\bar x(0)$ (for $\alpha = 200$ and the initial conditions
shown in Fig. \ref{fig:Fig4}b). It is intuitively clear that a slower producer growth would yield a smaller
increase. As the figure shows, we can find a region of extreme reduction ($s > 0.4$, which is
unlikely to appear in natural systems), which cannot be offset by the benefit from the public
good, thus producing no increase whatsoever in producer fraction. For lower values (roughly
between $0.1$ and $0.4$), $\Delta\bar x$ is positive, and increases as $s$ is lowered. Finally, for low $s$ (below
$0.1$), $\Delta\bar x$ increases further, reaching values comparable with $\bar x(0)$, implying that the global
producer fraction $\bar x$ almost doubles during growth, albeit transiently. The specific values of s
at which different results occur depend on the choice of $\alpha$ and of the $x_0$ distribution.
Nevertheless, the qualitative behavior of $\Delta\bar x$ remains the same.

The model thus provides insights into this public-good-mediated social interaction,
and implies that it leads to a transient, but potentially very significant, increase in producer
fraction. In the following section, we show that these predictions are in good agreement with
experiments on competitive growth of mixed populations of producers and non-producers.

\subsection{Comparison between experimental and theoretical results}
\begin{figure}[htb]
\centerline{
\includegraphics[width=\linewidth]{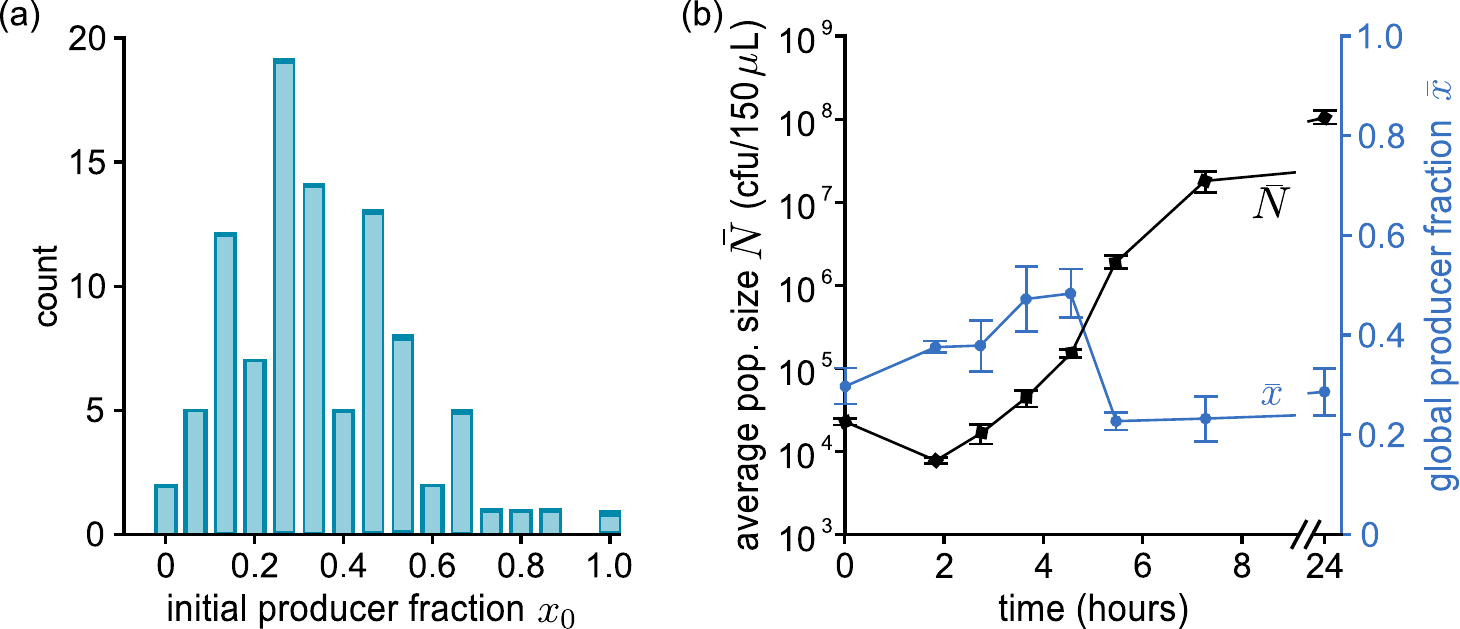}
}
\caption{Experimental results for the growth of a mixed metapopulation. (a) Sample distribution of
initial producer fractions in a 96-well plate. (b) Time course of the development of the total cell
number $\bar N(t)$ and global producer fraction $\bar x(t)$ for a metapopulation grown under extreme iron
limitation (KB/1 mM DP) in a 96-well plate shaken at $30^\circ$C. At given time intervals, samples are taken
from the wells, merged: $\bar N(t)$ is determined by counting cfu and $\bar x(t)$ is assessed based on the (green)
color of colonies. After a lag phase, populations begin to grow exponentially. During this phase, the
global producer fraction transiently increases, dips sharply, then stabilizes to its final value.
}
\label{fig:Fig5}
\end{figure}

We grew mixed populations composed of producers KP1 and non-producers 3E2
under extreme iron limitation (KB/1 mM DP), in which PVD is indispensable for iron uptake
and growth (see Supplementary Fig. S1). The metapopulation consisted of a 96-well plate (so
the metapopulation size is $M = 96$), and each well was inoculated with about $10^4$ cells.
Producers and non-producers in each well were mixed in stochastic proportions, sampled
from the distributions shown in Figs. \ref{fig:Fig4}b and \ref{fig:Fig5}a, which were derived from the weighted
average of three dice rolls (see Materials and Methods). These initial conditions mimic the
characteristic variability of small populations. The mean initial producer fraction was
$\bar x(0) \simeq 0.33$. As outlined in Fig. \ref{fig:Fig1}b, at given time points $t$, samples were taken from each
well, merged, and their average cell number $\bar N(t)=\frac{1}{M}\sum_i^M c_i +f_i$ and mean global producer
fraction $\bar x(t)$ were determined. Fig. 5 shows the results of these measurements (averaged over
four independent runs). On average, populations start growing after a lag phase of about 2 h
and enter stationary phase after around 8 h. The global producer fraction x initially increases,
up to a maximum $\bar x_\textnormal{max}\simeq 0.5$. After sharply dipping to $\bar x_\textnormal{min}\simeq 0.2$, it levels off to values
around $0.2-0.3$, and remains constant for at least 24 h. These results qualitatively agree with
those obtained by solving equations (4) for an analogous metapopulation (see previous section
and Fig. 4).

The only qualitative departure from the simulation results is that $\bar x$ drops towards the
end of growth phase ($t\simeq 6$h) in the experiments. Notably, however, this also corresponds to
an acceleration in population growth. Most probably, this stems from a change in the
metabolic state of cells, which is not captured by the simplified description encoded in the
equations (\ref{eq:4}).

We can also directly compare theoretical and experimental results. As initial
conditions for the simulations, we sample the values of $x_0$ from the same distribution as in the
experiments, and set $\bar n_0 = 10^{-3}$, which we estimated by dividing the mean minimum size
from experiments (taken at the end of the lag phase, so as to eliminate the slight population
decay) by the final yield. To set $s$, we considered that KP1 grows at a rate that is between
1.03 and 1.10 times lower than that for 3E2 (as determined previously), which corresponds to
a range for $s$ of between 0.03 and 0.09. Since the rate of approach to saturation $\alpha$ reflects
several complex processes, we opted to fit it.

\begin{figure}[htb]
\centerline{
\includegraphics[width=\linewidth]{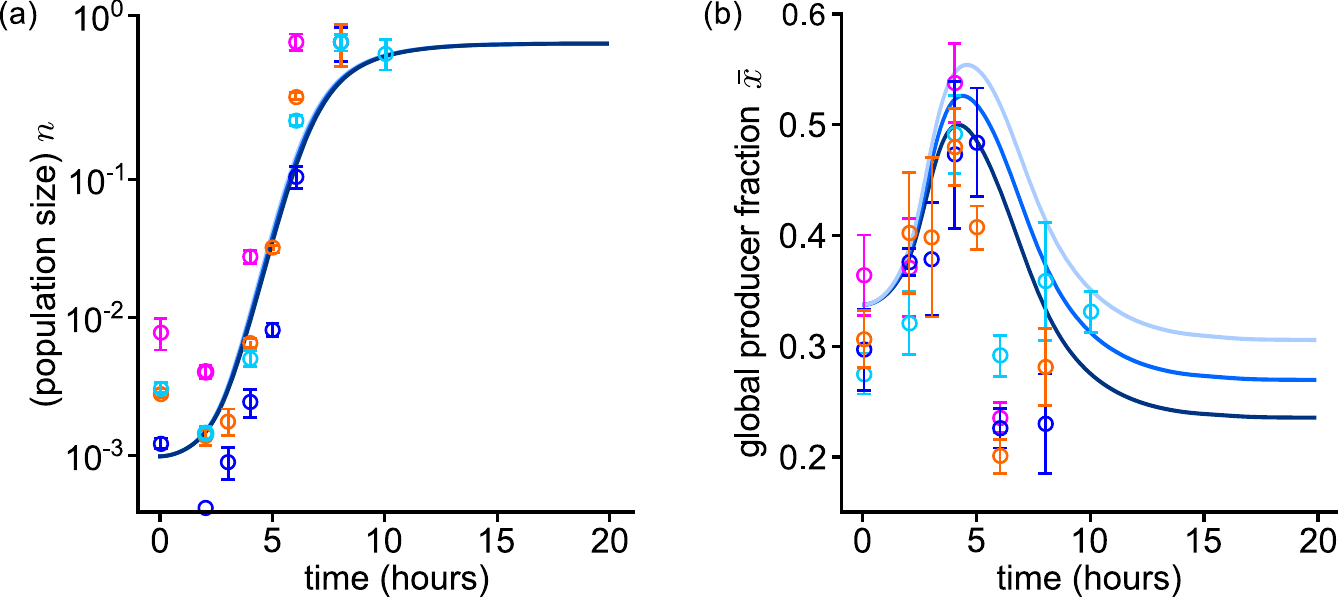}
}
\caption{Comparison of simulation and experimental results for the population size (a) and global
producer fraction (b) in a metapopulation. Solid lines represent numerical solutions of equations (\ref{eq:4})
for different values of the production cost $s$, and in (b) darker shades indicate higher values
($s\in\{0.03,0.05,0.07\}$). Dots of different colors indicate the results of different independent
experimental runs. Error bars are the result of three to five determinations of the respective parameter
at the given time point in one experiment. The population size $n$ is rescaled to the final yield (or
carrying capacity). The stochastic initial compositions are sampled from the distribution in Fig. \ref{fig:Fig4}b.
For appropriate values of the parameters (determined from fitting of the growth curve in (a)),
theoretical and experimental result agree.
}
\label{fig:Fig6}
\end{figure}

The data from four separate experiments (colored dots) and simulations for three
possible values of s and an appropriate saturation rate, $\alpha = 200$ (solid lines) are shown in
Fig. \ref{fig:Fig6}. To meaningfully compare the two sets of data, we also need to fix the global time scale
of simulations, which is done by fitting the slope of the exponential phase in Fig. \ref{fig:Fig6}a. The
increase in the global producer fraction observed in simulations agrees very well with
experiment (Fig. \ref{fig:Fig6}b): $\bar x$ grows to a maximum $\bar x_\textnormal{max} \simeq 0.5$ over similar periods, then decreases, and stabilizes to similar values.

Besides the aforementioned end-of-growth discrepancy – which seems to be due to
behaviors well beyond the scope of our simplified mathematical description – experimental
and theoretical results match.

\section{Discussion}
In this work, we showed that social interactions mediated by a public good result in a
transient increase in the global fraction of producers in a growing bacterial metapopulation.
By combining theoretical modeling and experiments, we were able to quantitatively describe
an exchange interaction involving a public good in a bacterial metapopulation.

We selected as our model system the native production of the iron-chelating
siderophore pyoverdine (PVD) in \textit{P. putida} KT2440 under iron limitation \cite{41,47}. We
characterized a constitutive producer (KP1) and a non-producer strain (3E2), and determined
the growth rate reduction due to producing PVD. Under the chosen conditions, PVD is
essential for iron acquisition and growth. We demonstrated that populations that produce
more PVD grow faster than those with less (under otherwise identical conditions), though the
magnitude of the benefit progressively diminishes as PVD accumulates, and eventually
vanishes when the available iron ceases to limit growth. Based on these experimental facts,
we constructed a set of differential equations that describes the growth of mixed populations
of PVD producers and non-producers. Solving these equations for a large metapopulation, we
found that, at first, the more producers (and thus more PVD) are present in the sub-
populations, the faster they grow. This generates a positive covariance between composition
and growth rate, which drives the global producer fraction up, in accordance with the Price
equation \cite{15,17,58}. As PVD accumulates, however, the benefit to cells eventually saturates,
reducing the advantage enjoyed by these producer-rich populations; meanwhile, populations
containing fewer producers begin to grow and ultimately catch up with the initially faster
ones. Therefore, the increase in the global fraction of producers is transient, both in
simulations and in experiments.

Previous experimental studies related similar phenomena to the so-called Simpson's
paradox \cite{11,12}. However, they considered an artificial bacterial system, in which both the need
for the public good and its production mechanism had been designed specifically for the
experiments. In contrast, we employed a native system and quantified its social interactions,
particularly the function and biochemical properties of the public good. Our analysis also
shows that, without mechanisms to sustain it, the Simpson-related increase can only be
transient. This conclusion is also compatible with previous qualitative predictions \cite{13,14,15,16}, based
on game theory models with implicit public goods. However, in contrast to our experiments,
these studies predict that the producer fraction should peak at the end, instead of the mid-point
of exponential growth. This indicates that simple cost-benefit considerations do not suffice to
describe the social interaction. Inclusive fitness models have been used to describe an
analogous scenario in wild-type \textit{P. aeruginosa}, reaching qualitative conclusions compatible
with our results \cite{29,30}. Similarly to game-theoretical approaches, however, they remain mainly
conceptual \cite{59}. Our systems biology approach, instead, provides a simple description, with
testable quantitative predictions, as well as important insights into the social interaction.
In metapopulation settings, diffusion, dispersal, and mobility affect public good
interactions \cite{30,60}. Besides these factors, our results highlight the potential role of the timing of
dispersal. Some studies already pointed to dispersal timing, by considering a metapopulation
that periodically splits into groups and merging these again to reform the pool. After several
cycles, the metapopulation might develop stable coexistence of the strains \cite{13,14,15,16}, or even have
the producers fixate \cite{11,12}. Testing this process, however, requires Poisson dilution conditions
which implicate very low initial densities of producer cells. As a consequence, large fractions
of cells die under iron-limiting conditions before physiologically effective PVD
concentrations are reached. Therefore, a repetitive scenario of group formation and merging is
experimentally not feasible for our well mixed cultivation conditions. In principle,
introduction of a non-selective growth phase may rescue such a scenario \cite{61}.

An interesting next step will be to include regulatory aspects in our system. Like many
other bacteria, the wild-type \textit{P. putida} KT2440 continually senses changes in environmental
conditions, and uses this information to tune production of the public good \cite{62,63,64}. By employing
constitutive producer strains, we shifted the focus more on the social role of PVD itself, while
replicating a potential earlier stage of evolution (if PVD production evolved before
regulation). Our model also indicates that a cost-saving strategy such as down-regulation of
PVD production as a consequence of PVD accumulation is not sufficient to prevent the long-
term decline of the global fraction of producers, because all populations with producers
eventually accumulate the same PVD concentration. So accounting for regulation, which has
been shown to also impact growth \cite{65}, will also necessarily involve elaborate production
curves \cite{63} and cost-saving strategies \cite{66}. Ultimately, adaptive production raises complex
questions about how cells shape the ecological and environmental conditions in which they
interact \cite{67}.
Another possible extension would be to allow privatized use of the public good.
Privatized use of siderophores, in particular, has been shown to introduce fascinating social
dynamics into intra- and inter-species competition \cite{35,68,69,70}. Limited diffusion and private use
have important social consequences \cite{32,33}. Indeed, several studies have intensely debated under
what conditions the secreted siderophores actually behave as public goods \cite{42,71,72,73}. In our
conditions, however, populations seem to behave as well-mixed, with negligible privatization.
Taken together, our work uses a simplified setting to highlight the determinant role of
public goods in social interactions and population dynamics. For example, we showed the
profound consequences of the public good's accumulation and saturating benefits, which
simple game-theoretical considerations would fail to describe. Our approach could clearly be
extended to investigate the fundamental principles underlying different interactions and
bacterial systems. Thereby it should stimulate more mechanistic analyses of bacterial social
interactions and their impact on population development.

\section*{Acknowledgements}
This work was supported by the Deutsche Forschungsgemeinschaft through grant SPP1617
(JU 333/5-1, 2 and FR 850/11-1, 2). We thank Michelle Eder (HJ lab) for excellent technical
assistance. P. putida strain 3E2 was kindly provided by P. Cornelis (Vrije Universiteit
Brussels, Belgium).

\section*{Author Contributions}
Designed and performed the experiments: FB, HJ.; designed and performed theoretical
analysis: KW, ML, EF; analyzed the experimental and computational data: FB, KW, ML, EF,
HJ. Wrote the paper: KW, ML, EF, FB, HJ.

\section*{Materials and Methods}
\subsection*{Strains and growth conditions}
\textit{Escherichia coli} DH5$\alpha$ [F-$\varphi$80d lacZ $\Delta$M15 Δ(lacZYA-argF) U169 deoR recA1
endA1 hsd R17(rk-,mk+) phoA supE44 $\lambda$- thi-1 gyrA96 relA1] was used as the carrier for
plasmids. \textit{Pseudomonas putida} KT2440 and the derived strain 3E2 (non-producer) \cite{47} 47 were
employed as PVD producer (wild-type) and non-producer, respectively. \textit{E. coli} strains were
grown in lysogeny broth (LB) at 37°C, and \textit{P. putida} strains were grown at $30^\circ$C in King’s
medium B (KB) \cite{74}. KB medium was supplemented with 100 $\mu$M FeCl$_3$ and/or 1 mM of the
iron chelator 2,2’-dipyridyl (DP) where indicated. Solid media were LB or KB with 1.5\%
agar.

\subsection*{Generation of the constitutive PVD producer strain KP1}
A \textit{P. putida} strain that constitutively produces PVD was generated by placing a copy
of the pfrI gene under the control of the constitutive promoter P$_{A1/04/03}$ \cite{54}. For this purpose,
P$_{A1/04/03}$ and the \textit{pfrI} gene were amplified by PCR from the plasmid mini\textit{Tn7}(Gm)P$_{A1/04/03}$ ecfp-
a \cite{75} and the \textit{P. putida} genome, respectively, cloned into plasmid pUC18R6K-mini-\textit{Tn7}T-Gm,
and inserted at the \textit{att}Tn7 site in \textit{P. putida} KT2440 following a mini-\textit{Tn7} protocol for
\textit{Pseudomonas} \cite{76}. The resulting \textit{P. putida} strain KP1 was verified by PCR amplification of
corresponding genome regions and sequencing. All oligonucleotide primers used for strain
generations and verification are listed in Supplementary Table S2.

\subsection*{Quantitative analysis of PVD production}
Pre-cultures of the respective strains were grown in iron-replete medium (KB/200 $\mu$M
FeCl$_3$) at 30$^\circ$C for 18 h. The pre-cultures were used to inoculate the appropriate media for the
growth of the cultures used in experiments ($N_0 = 10^7$ cells mL$^{-1}$). Experiments were
performed in 24-well plates (2 mL culture/well). The plates were shaken at 300 rpm at 30$^\circ$C.
At given time points samples were taken and the optical density at 600 nm was measured.
Subsequently, cells were removed by centrifugation, and the relative PVD content was
determined by measuring the fluorescence emission at 460 nm (excitation 400 nm). PVD
production was analyzed under iron limitation (KB/1 mM DP; KB/100 μM FeCl$_3$/1 mM DP)
and iron replete conditions (KB; KB/100 $\mu$M FeCl$_3$). Each individual experiment was
performed with three parallel replicates. A minimum of three independent experiments were
conducted per condition.

\subsection*{Growth characteristics of strains under different environmental conditions}
Pre-cultures of the respective strains were grown in iron-replete medium as described
above for the analysis of PVD production, and used to inoculate the appropriate media for
growth of the cultures used in experiments ($N_0 = 10^7$ cells mL$^{-1}$). Experiments were
performed in 96-well plates (150 $\mu$L culture/well). The plates were shaken at 300 rpm at
30$^\circ$C. Growth was followed by measuring the optical density at 600 nm using a microplate
reader (Infinite \textregistered M200 from Tecan Trading AG). The measurement was controlled and
monitored with the i-control$^\textnormal{TM}$ Software from Tecan Trading AG (30°C, shaking at 280 rpm,
880 s per cycle, minimum 80 cycles). Each condition was implemented in six replicates per
experiment, including medium blanks. For low cell numbers (e.g., $N_0= 10 ^4$ cells 150 $\mu$L$^{-1}$),
growth was analyzed by determining colony forming units (cfu) over time (threefold per time
point). The specific growth rate $\mu$ represents a quantitative measure of growth in the
exponential phase and was calculated using the following equation: $\mu=\frac{\ln\left(N(t_2)-N(t_1)\right)}{t_2-t_1}$
.

\subsection*{Quantitative assessment of the benefit of PVD}
The benefit conferred by PVD was quantified under iron-limiting conditions (KB/1
mM DP) with the non-producer strain 3E2. PVD was isolated according to a previously
described protocol \cite{77} and added to the medium at concentrations of between 0 and 20 $\mu$M.
Growth was monitored via optical density measurement, and the specific growth rate μ was
calculated as described in the previous paragraph. Each individual experiment was performed
with four parallel repeats per PVD concentration, and three independent experiments of this
type were conducted per PVD concentration.

\subsection*{Determination of PVD sharing in mixed culture}
Cells were grown in KB/1 mM DP (initial producer frequency $\bar x(0)\simeq 0.33$, $N_0 = 10^4$
cells/150 $\mu$L, 96-well plate format) at 30$^\circ$C. Colony forming units (cfu) were determined at
given time points (five replicates per time point), and producer and non-producer cells were
discriminated by colony color and size. Three independent experiments were performed, each
yielding similar results.

\subsection*{Stability of PVD in KB medium with and without bacteria}
Medium without cells and medium containing about $10^7$ cells mL$^{-1}$ of the non-producer were supplemented with 2$\mu$M PVD and incubated at 30$^\circ$C for 48 h. At given time
points samples were taken, and the relative PVD contents of medium and of the supernatant
of medium with cells were determined by measuring the fluorescence emission of PVD at 460
nm (excitation 400 nm).

\subsection*{Competitive growth experiments}
To analyze the impact of the initial producer frequencies $x_0$ on growth, strains KP1 and
3E2 were mixed in KB/1 mM DP (96-well plate, $N_0 = 10^4$ cells 150 $\mu$L$^{-1}$,
$x_0\in\{0, 0.1, 0.2, 0.3, 0.5, 0.75, 1.0\}$). Total cell numbers were determined at the end of the lag
phase and after 8 h of shaking at 30$^\circ$C by counting cfu. For each condition, a minimum of
three individual experiments were performed. To analyze the development of the total cell
number $\bar N (t)$ and global producer frequency $\bar x (t)$ in metapopulations, a random distribution
of the initial producer frequency $x_0$ was established by rolling three dices. The values of each
triplet were weighted (lowest 2/3, middle 2/9 and highest 1/9) and rounded to yield sixteen
different values from 0 to 15 that are equivalent to sixteen different initial producer
frequencies $x_0$ ranging from $0$ to $1.0$ and result in an initial average global cooperator fraction
$\bar x (0)$ of about $0.33$. Cells were grown in KB/1 mM DP (96-well plate, $N_0= 10^4$ cells 150 μL $^{-
1}$) at 30$^\circ$C while shaking at 300 rpm. At given time points aliquots of each well were merged
and $\bar N (t)$ was determined by counting cfu. The global producer frequency $\bar x(t)$ was obtained
based on differences in the color and size of the colonies of KP1 and 3E2 on KB agar plates
(minimum three replicates per time point).

\subsection*{Stochastic initial conditions and ensemble averages}
We generate all triplets of the integers between 0 and 5 to simulate the results of a
sequence of (simultaneous) throws of three dice. Since the weights in the experimental
procedure are assigned based on the order of the rolled values, we order the “rolled” values
within each triplet from lowest to highest. This results in a table of all possible 3-dice rolls,
which we can use directly to generate the initial conditions and simulate equations (4). To
speed up calculations, however, we remove duplicate triplets: for example, 113, 131, 311 are
different triplets before sorting, but are the same after. Once we remove the duplicate
combinations, we assign the appropriate probability to them, i.e. the number of ways to
produce them before sorting divided by the total number of triplets. With a minimal
combinatorics, one can compute the total number of triplets (6 ! = 216), and the multiplicities
of triplets: those with three equal values have only one way to appear before sorting; those
with two equal values have three; those with all different values have six.

To simulate the time evolution of $n$ and $x$, we generate the initial composition $x_0$ for
each triplet, using the weighted average described above. After setting $n_0$ , $\alpha$, and $s$, the
temporal evolution of the average $\bar x$ and $\bar n$ in ensembles of populations can be computed by
solving equations (4) for each of the allowed values of $x_0$ and weighting it using the relative
probability, computed as described above.

\renewcommand{\thefootnote}{\fnsymbol{footnote}}
\renewcommand{\thefigure}{S\arabic{figure}}
\renewcommand{\thetable}{S\arabic{table}}
\renewcommand{\theequation}{S\arabic{equation}}
\setcounter{figure}{0}
\setcounter{equation}{0}

\newpage

\begin{widetext}
\begin{center}
\LARGE Non-selective evolution of growing populations\\Supplementary Information
\medskip
\setcounter{page}{1}

\large{Karl Wienand, Matthias Lechner, Felix Becker, Heinrich Jung, and Erwin Frey}
\end{center}

\section*{Derivation of the growth rate $\mu(p)$}
Assume a cell is born with an internal iron concentration $Fe_\textnormal{in}(0)$ and a volume $V(0)$. Let $p$ be
the concentration of PVD-Fe complexes (which we take to be the same as that of PVD; see
text). Each cell, then, incorporates iron ions at a constant rate $k~p$ that is proportional to the
concentration of PVD. So at time $t$ after its birth, the cell has accumulated $kpt$ iron ions. Its
internal iron concentration $Fe_\textnormal{in}(t)$, then, is $Fe_\textnormal{in}(0)V(0)$ (that is, the number of iron atoms at
birth), plus the iron it has collected, all divided by the volume birth $V(t)$ it has reached:
\begin{equation*}
Fe_\textnormal{in}(t)\frac{Fe_\textnormal{in}(0)V(0)+kpt}{V(t)}\,.
\end{equation*}
Because cells try to maintain iron concentration homeostasis, we can consider Fe in to be
constant. Moreover, as long as iron is the limiting factor for growth, the growth rate $\mu(p)$
depends only on the PVD concentration.

On average, cells divide at time $t_D = 1/\mu(p)$, given that growth is logistic. Moreover, at
the moment of division, the cell has attained twice the volume of its future daughters:
$V(t_D) = 2V(0)$. With these substitutions in the above equation, and minimal algebra, we
obtain equation (\ref{eq:2}).

\section*{Estimation of iron incorporated into cells}
The iron content of a bacterial cell ranges from$\sim 10^5$ to $10^6$ atoms per cell \cite{S1,S2}. In our
experimental setup, cells reach a maximum density of $2\times10^7$ cells per 150 $\mu$L at the end of
the exponential growth phase accumulating in total $2\times10^{12}$ to $2\times10^{13}$ iron atoms. We
determined the iron concentration of our KB preparation by atomic absorption spectroscopy
and found a concentration of $\sim8~\mu$M (corresponding to $\sim7.2\times10^{14}$ iron atoms per 150 $\mu$L KB
medium). Using these numbers we calculated that $\sim 0.28$ to 2.8\% of the total iron of KB is
incorporated into cells by the end of the exponential growth phase.

\section*{Impact of simple regulation}
Our conclusions are robust (at least qualitatively) against simple cost-saving strategies.
More specifically, PVD accumulation makes it so that stopping synthesis to save costs is not
sufficient to prevent the long-term decline of the global fraction of producers, as we prove in
the following.

Let the synthesis rate $\sigma$ and the production cost $s$ in equation (\ref{eq:3}) depend on the
pyoverdine concentration $p$. As a general form of cost-saving regulation, we consider that
producers cease to synthesize PVD when its concentration reaches some threshold $p_c:\sigma(p_c) = 0$. After synthesis stops, there producers incur no further cost and thus grow at the
same rate as non-producers: $s(p_c) = 0$. If a population has any amount of producers (that is,
if $x_0\neq 0$), at least some PVD is produced and, because it accumulates, its concentration
steadily grows until it reaches $p_c$. In the long run, then, all populations with $x_0\neq0$ end up
with the same PVD concentration (namely $p_c$).

As in our analysis, producer-rich populations grow faster for a limited time, then less
producing ones catch up. Concurrently, the producer fraction within each population declines
(during production) or stays constant (when production stops). This tension engenders a
"Simpson's paradox" setting much like the one we presented in the case of constitutive
production. Furthermore, because all populations reach the same ultimate size, the final value
of $\bar x$ is simply the average of the producer fraction $x$ in each population. These values are
lower than the respective $x_0$, because of the cost of production.

\begin{figure}
\centerline{
\includegraphics[width=0.4\linewidth]{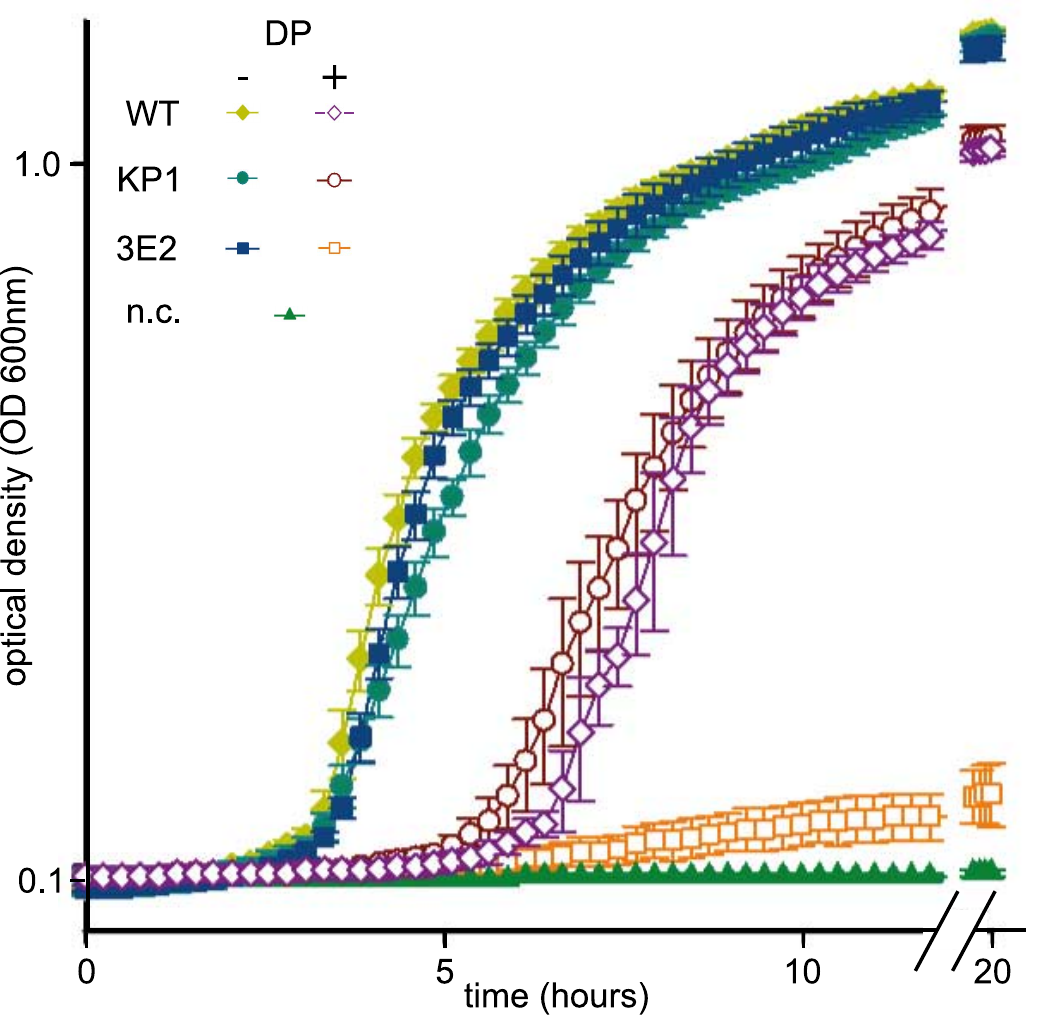}
}
\caption{
Growth of \textit{P. putida} KT2440 and the derived strains KP1 and 3E2 under iron
replete and limiting conditions. Cells were grown in KB supplemented with iron (KB/100
$\mu$M FeCl$_3$, full symbols) and in KB with iron chelator DP (KB/100 $\mu$M FeCl$_3$/1 mM DP,
empty symbols). When iron is more available, PVD is not needed for growth, 3E2 and WT
grow about at the same rate, and KP1 grows slower. Under iron limitation, producing strains
(KP1 and WT) benefit from production and grow much faster than non-producing 3E2. The
experiment was performed as described in the legend of Fig.\ref{fig:Fig3}a.
}
\label{fig:FigS1}
\end{figure}

\begin{figure}
\centerline{
\includegraphics[width=0.8\linewidth]{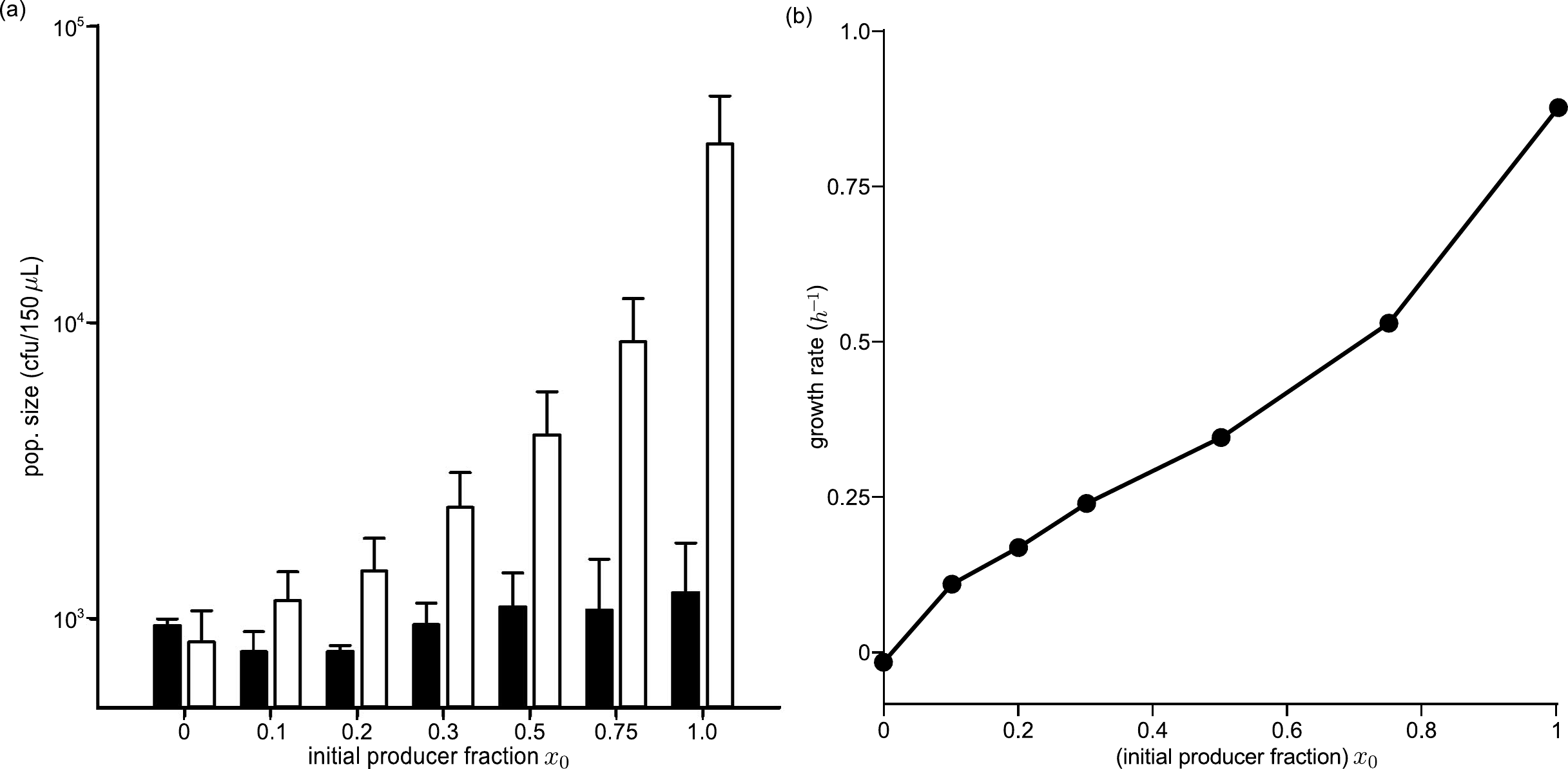}
}
\caption{
Impact of the initial producer fraction $x_0$ on the growth of mixed populations
under iron limiting conditions. \textbf{(a)} Impact of the initial producer fraction $x_0$ on the growth
yield. Strains KP1 and 3E2 were grown in mixed culture under iron-limiting conditions (KB/1
mM DP, $N_0$ about $10^3$ cells/150 $\mu$L, 96-well plate format) with the given initial producer
frequencies $x_0$. Total cell numbers were determined by counting cfu at the beginning of the
experiment (black columns) and after 8 h of incubation (weight columns). For each condition,
minimum three individual experiments were performed. \textbf{(b)} Impact of the initial producer
fraction $x_0$ on the specific growth rate μ. Mixed cultures of strains KP1 and 3E2 with $x_0$
values between 0 (=100\% 3E2) and 1 (=100\% KP1) were incubated in shaking 96-well
microtiter plates at 30$^\circ$C ($N_0 = 10^7$ cells mL$^{-1}$). Growth was analyzed by measuring the
optical density at 600 nm using a Tecan microplate reader. $\mu$ was determined for each
condition from the exponential phase of the resulting growth curves. All growth parameters
represent the means of five growth experiments. Deviations were <10\% of the mean value.}
\label{fig:FigS2}
\end{figure}

\begin{figure}
\centerline{
\includegraphics[width=0.9\linewidth]{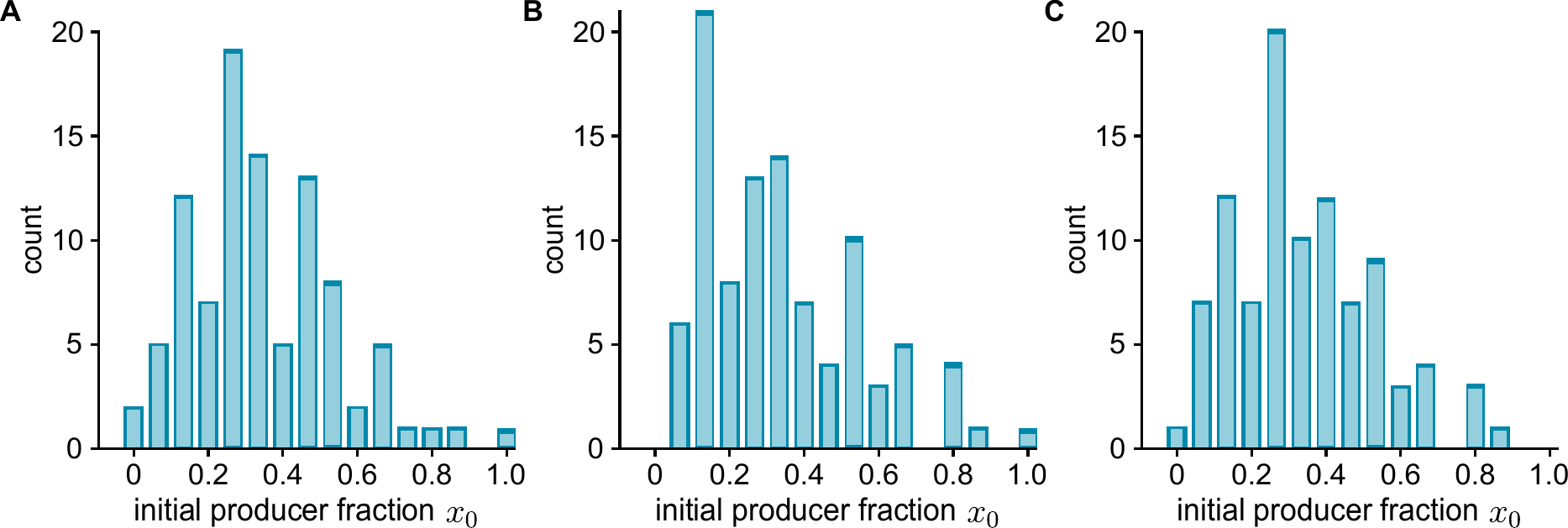}
}
\caption{
Distribution of initial compositions $x_0$ in three replicates of the experiment.
Although specific values differ, the overall features of the distribution remains the same. Most
populations start mixed, with $x_0$ between 0.1 and 0.5. When at all present, populations with
all producers or no producers are very rare.
}
\label{fig:FigS3}

\end{figure}

\begin{table}
\begin{tabular}{|c|c|c|c|c|}
\hline \rule[-2ex]{0pt}{5.5ex} \textbf{Growth
medium} & \textbf{$\mathbf{\mu_{WT}}$
(h$^{-1})$} & \textbf{$\mathbf{\mu_{3E2}}$
(h$^{-1})$} & \textbf{$\mathbf{\mu_{KP1}}$
(h$^{-1})$} & $\mathbf{\mu_{3E2}/\mu_{KP1}}$ \\ 
\hline \rule[-2ex]{0pt}{5.5ex} KB$^a$ & 0.797 & 0.776 & 0.736 & 1.054 \\ 
\hline \rule[-2ex]{0pt}{5.5ex} KB/100 $\mu$M FeCl$_3~^a$ & 0.766 & 0.749 & 0.692 & 1.082 \\ 
\hline \rule[-2ex]{0pt}{5.5ex} KB/DP$^a$ & 0.489 & 0.004 & 0.588 & 0.007 \\ 
\hline \rule[-2ex]{0pt}{5.5ex} KB/100 $\mu$M FeCl$_3$$^a$ & 0.657 & 0.030 & 0.610 & 0.049 \\ 
\hline \rule[-2ex]{0pt}{5.5ex} KB/100 $\mu$M FeCl$_3~^b$ & 1.237 & 1.238 & 1.201 & 1.031 \\ 
\hline \rule[-2ex]{0pt}{5.5ex} KB/100 $\mu$M FeCl$_3~^c$ & n.d. & 1.182 & 1.077 & 1.097 \\ 
\hline 
\end{tabular} 
\caption{
Specific growth rate of P. putida KT2440 (WT), the non-producer
(3E2), and the constitutive PVD producer (KP1) under iron-rich and iron-
limiting conditions.\\
$^a$ The specific growth rate $\mu$ was calculated from the growth curves shown in Fig \ref{fig:Fig1}(c). Cells
were grown in shaking 96-well microtiter plates at 30$^\circ$C ($N_0= 10^7$ cells mL$^{-1}$). Growth was
analyzed by measuring the optical density at 600 nm using a Tecan microplate reader.\\
$^b$ Cells were grown in shaking 24-well microtiter plates at 30$^\circ$C ($N_0 = 10^7$ cells mL$^{-1}$). Every
hour 50\% of the culture was replaced with fresh medium. Growth was analyzed by measuring
the optical density at 600 nm using a 1-mL cuvette (d=1 cm).\\
$^c$ Cells were grown in shaking 96-well microtiter plates at 30$^\circ$C ($N_0 = 10^4$ cells mL$^{-1}$). Growth
was analyzed by determination of colony forming units, \textit{cfu}.\\
All growth parameters represent the mean of five to fifteen growth experiments. Deviations
were <10\% of the mean value.}
\end{table}

\begin{table}
\begin{tabular}{ll}
\textbf{Name} & \textbf{Sequence (5'...3')} \\ \hline 
\textbf{Generation of P. putida KP1} &  \\ 
$P_{A1_04_03}$ bw kpn & AAATAGGGGGGTACCCGCACATTTCCC \\ 
$P_{A1_04_03}$ mod2 & TTCCGCCATGCTTAATTTCTCCTCTTT \\ 
\textit{pfrI} start mod2 &AAATTAAGCATGGCGGAACAACTATCC \\ 
\textit{pfrI} end mod2 & TGCGGCGTTGGATCCGCTGCGAGTTATTGGCCG \\ 
\textbf{Sequencing insert on plasmid} &  \\ 
mini Tn7 reverse MCS & TTGCATTACAGTTTACGAACCGAAC \\ 
\textbf{Sequencing Tn7 insertion on genome} &  \\ 
checkdown primer trans & GTCTTATTACGTGGCCGTGC \\ 
Primer TN7R as & CCACGCCCCTCTTTAATACG \\ 
tn7left s & TTTGTCATTTTTAATTTTCG \\ 
checkup primer trans & GCAGGAGCCGATGAGACAGA \\
\end{tabular}
\caption{Oligonucleotides used in this investigation}
\end{table}

\subsection*{Movie: Temporal evolution of $\bar x$ in a metapopulation.} The evolution of $\bar x$ was obtained by
solving equations (\ref{eq:4}) together with the evolution of the joint distribution of sizes $n_i$ and
compositions $x_i$.\\
\url{https://static-content.springer.com/esm/art\%3A10.1038\%2Fs41598-018-22306-9/MediaObjects/41598_2018_22306_MOESM2_ESM.mp4}

\end{widetext}
\end{document}